\documentstyle[aps,prb,multicol,epsf]{revtex}

\begin{document}

\draft
\title{Magic Islands and Barriers to Attachment: A~Si/Si(111)7$\times$7
Growth Model}
\author{J. Myslive\v{c}ek\cite{byline1} and T. Jarol\'{\i}mek}
\address{Department of Electronics and Vacuum Physics, Faculty of
Mathematics and Physics, Charles University, V~Hole\v{s}ovi\v{c}k\'ach~2,
180~00~Praha~8, Czech Republic}
\author{P. \v{S}milauer}
\address{Institute of Physics, Academy of Sciences of the Czech Republic,
Cukrovarnick\'{a}~10, 162~53~Praha~6, Czech Republic}
\author{B. Voigtl\"ander and M. K\"astner}
\address{Institut f\"ur Grenzfl\"achenforschung und Vakuumphysik,
Forschungszentrum J\"ulich, 52425~J\"ulich, Germany}
\date{\today}
\maketitle
\begin{abstract}
Surface reconstructions can drastically modify growth kinetics
during initial stages of epitaxial growth as well as during the process
of surface equilibration after termination of growth. We investigate the
effect of activation barriers hindering attachment of material
to existing islands on the density and size distribution of
islands in a model of homoepitaxial growth on Si(111)7$\times$7
reconstructed surface. An unusual distribution of island sizes
peaked around ``magic'' sizes and a steep dependence of the
island density on the growth rate are observed.
``Magic'' islands (of a different shape as compared to those
obtained during growth) are observed also during surface
equilibration.
\end{abstract}

\begin{multicols}{2}

Investigation of island structures formed during the initial stages
of epitaxial growth allows us to explore kinetic mechanisms that
govern the ordering of deposited atoms.\cite{ven1}
A~lot of attention has recently focused on the time- and
growth-conditions dependence of the island density \cite{smil1}
as well as on the bell-shaped distribution of island sizes
\cite{smil1} whose origin can be traced back to the distribution of
island capture zones.\cite{voronoi} However, real growth systems
are invariably more complicated than the idealized models of
epitaxy commonly used. For example, the presence of
surface reconstructions can completely change the growth behavior.

In homoepitaxy of Si on Si(111)7$\times$7 reconstructed surface, a
process of ``reconstruction destruction'' was described by
Tochihara and Shimada.\cite{tochi1} The need to cancel surface
reconstruction around a growing island gives rise to
{\it barriers to attachment\/} of new material to existing
islands. Growth with barriers to attachment has been already
studied theoretically: The dependence of the island density
on growth conditions
was explored using analytic methods,\cite{kandel,thib}
while kinetic Monte Carlo (KMC) simulations of a simple growth
model revealed an island size distribution multiple-peaked
around ``magic'' sizes.\cite{magic}

Here we present a detailed KMC model of
Si/Si(111)7$\times$7 molecular beam epitaxy (MBE), with
barriers to attachment included.  With the help of this model
we investigate the time- and growth-rate dependence of the
island density, the shape of the island-size distribution,
as well as island decay and filling of
vacancy islands on the surface. The results of our simulations
compare favorably to available experimental data about the
Si/Si(111)7$\times$7 system. We also discuss those features
of the model kinetics that are specific to growth with barriers
to attachment.

Dynamics of Si/Si(111)7$\times$7 MBE growth was experimentally
studied
by Voigtl\"ander {\it et~al.\/}\cite{bert1,bert2,xcomp2} The
most interesting feature observed, the existence of kinetically
stabilized magic sizes in the island size distribution, was
reported and KMC modeled in Ref.~\onlinecite{magic}.
The model discussed in this work is a generalization of the
model from Ref.~\onlinecite{magic} which is in turn based on the
``reconstruction destruction'' model of Si/Si(111)7$\times$7
growth proposed by Tochihara and Shimada.\cite{tochi1} The
material is deposited in units (cf. below) that are
randomly placed onto sites of a honeycomb lattice.
These sites represent half-unit cells (HUCs) of
Si(111)7$\times$7 surface reconstruction.
Two types of HUCs exist: In unfaulted HUCs (marked U), a surface
atom bilayer follows bulk bilayer stacking, in faulted HUCs
(marked F), the surface bilayer is 30$^{\circ}$ rotated with
respect to the bulk, forming a stacking fault.\cite{taka}
Material in a HUC may be either
non-transformed (marked $X$, models the free Si adatoms
diffusing on the reconstructed Si surface)
or transformed (marked $T$, corresponds to a HUC,
where the material underwent the reconstruction destruction
process and subsequent crystallization). The model thus becomes
{\it de facto\/} a two-species one. The non-transformed units
of the material randomly walk on the surface, meet each other
and transform.
Transformation is an activated process, activation energy for
transformation is supposed to be higher in F HUCs due to the need to
remove the stacking fault.\cite{tochi1}

In the experiment, deposited material does not diffuse as
HUCs.\cite{video} The use of units of deposited material allows us to
model easily collective processes during "reconstruction destruction"
around island edges. Since, in general, the processes at
island edges determine behavior of growth models (as they
determine growth behavior of real systems), the
simplification of material deposition
and surface diffusion should not affect the model behavior
in a substantial way.
Our simulation scheme is coarse-grained and ignores
all processes on the length scales smaller than a HUC and on
the time scales shorter than the time required to transport
material from one HUC to its nearest neighbor.

The hopping rate for a HUC hop is
$\nu_D$$=$$\nu_{0}\exp (-E_{D}/k_BT)$ where
$E_{D}$$=$$E_{S}+(x+t)E_N^X$ for
an X-HUC, $E_{D}$$=$$E_{S}+xE_N^X+tE_N^T$ for a T-HUC, $x$
and $t$ being the numbers of X and T neighbors, respectively,
$E_N^X$ bond strengths of $X-X$ and $X-T$ pairs,
$E_N^T$ bond strength of a $T-T$ pair, and
$E_{S}$ the surface barrier to diffusion. The rate of an
X-HUC transformation is
$\nu_T$$=$$\nu_{0}\exp(-E_T/k_BT)$ where
$E_T^F=E_{A}-tE_{\rm edge}$ for an F-HUC,
$E_T^U=E_{A}-tE_{\rm edge}-E_{\rm diff}$ for a U-HUC,
$E_{A}$ being the barrier to attachment, $E_{\rm edge}$ a
decrease in the barrier due to a transformed neighbor,
and $E_{\rm diff}$ the barrier difference for F- and U-HUC
overgrowth. Transformation of an island begins at an F-HUC
with $\geq$2 X-neighbors, and the rate of this nucleation
process is $\nu$$=$$\nu_{0}\exp[(E_{T}-E_{\rm edge})/k_BT]$.
The model has seven parameters, here we report
results for $\nu_{0}$=10$^{13}$~s$^{-1}$, $E_{S}$=1.5~eV,
$E_N^T$=0.3~eV, $E_N^X$=0.1~eV, $E_{A}$=2.3~eV,
$E_{\rm edge}$=$E_{\rm diff}$=0.35~eV, which gave the best
agreement with experimental
results. Using the model, we tried to reproduce both growth
and equilibration processes on the Si/Si(111)7$\times$7
surface on a real time- and spatial
scale.  The HUC in the model is thus considered to be
1~bilayer (BL) of Si(111) thick, of a triangular shape with the
edge length of $a$=26.9~{\AA} and consisting of 49 Si atoms.
All calculations were performed on 200$\times$200~HUC lattice with
periodic boundary conditions.

{\it Growth.\/} We tried to fit the exponent $\chi$ given by
\begin{equation}
\label{grexp}
N \approx F^{\chi}
\end{equation}
where the dependence of the island density $N$ on flux $F$ at a
constant temperature is measured.\cite{ven1,smil1,kandel}
The experimental value of
$\chi$=0.75 for $T$=680~K and $T$=770~K was reported in
Ref.~\onlinecite{xcomp1}. In this experiment, samples were prepared
at a given temperature by deposition of $\approx$0.15~BL Si on
the Si(111)7$\times$7 surface followed by rapid quenching to room
temperature.  The experimental morphologies of the layer can
therefore be considered snapshots from the Si/Si(111)7$\times$7
surface morphology evolution.

\begin{figure}
\epsfxsize=7.2cm
\centerline{\epsfbox{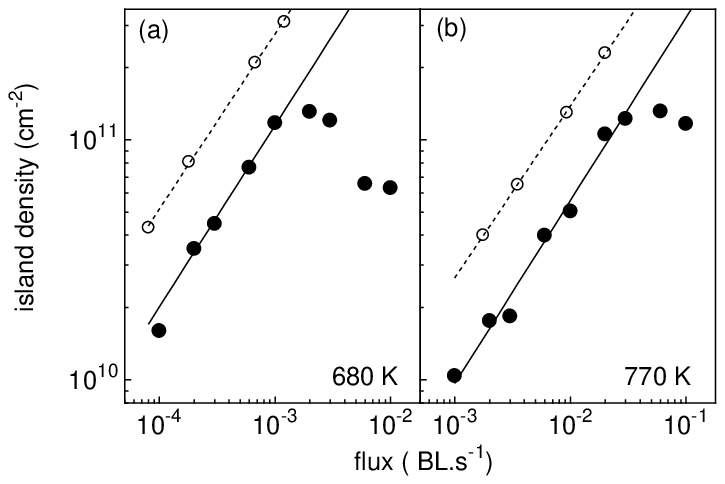}}
\caption{Flux dependence of the island density $N$ for
Si/Si(111)7$\times$7 MBE is $N$$=$$F^\chi$ with $\chi$$=$0.75.
Experimental ($\circ$) (Ref.~\protect\onlinecite{xcomp1})
and modeled ($\bullet$) $N(F)$ dependencies for 680 K~(a) and 770 K~(b)
are shown. In the model, $N$ at $\Theta_{\rm tot}$=0.15~BL was measured.}
\label{grexpg}
\end{figure}

In the model, we calculated the flux dependence of the density
of {\it transformed\/} (i.e., crystalline) islands $N$ at
constant {\it total\/} coverage $\Theta_{\rm tot}$
(Ref.~\onlinecite{com1}). Results are shown
in Fig.~\ref{grexpg}. At lower fluxes, a power-law
$N$$=$$F^\chi$ dependence with $\chi_{680}$=0.76$\pm$0.03,
$\chi_{770}$=0.75$\pm$0.04 is observed.
In the experiment, disordered growth occured
at high fluxes. In the model, deviations from the
power-law behavior of $N$$=$$N(F)$ are observed.

\begin{figure}
\epsfxsize=6.3cm
\centerline{\epsfbox{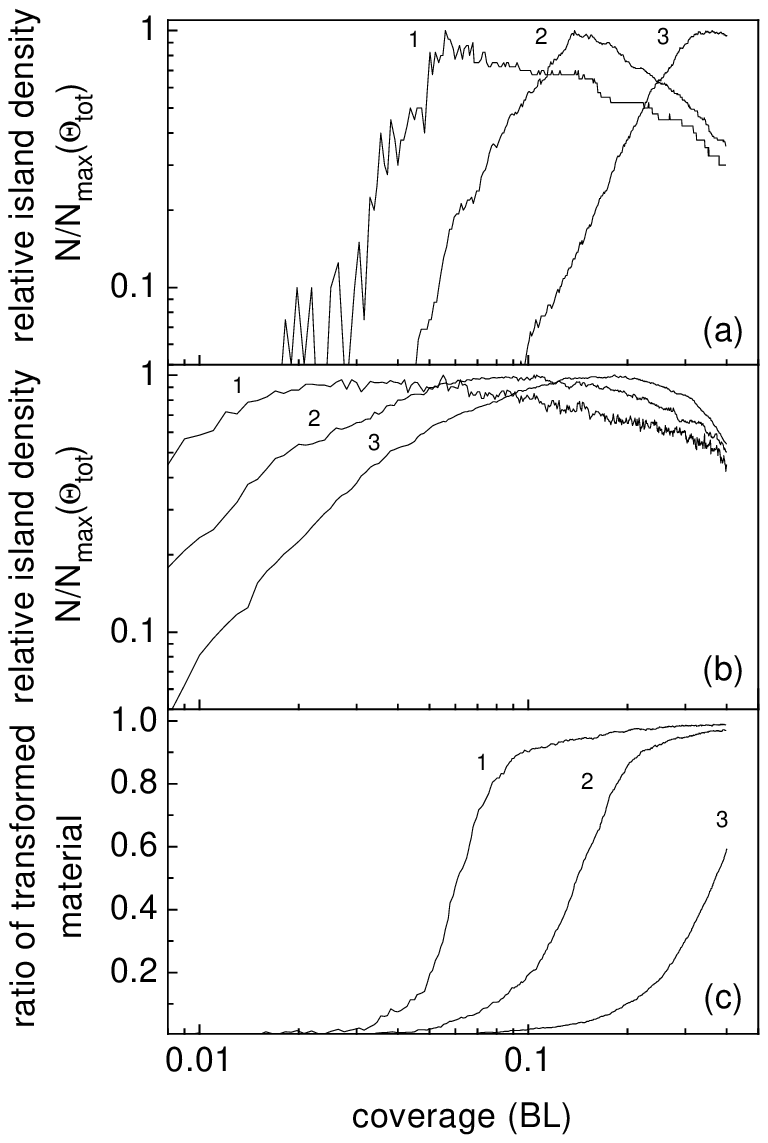}}
\caption{Evolution of the island density vs. total coverage for a
model with the barriers to attachment depends on flux (a) and
differs from that of a standard growth model (b),
Ref.~\protect\onlinecite{commin}.
The barrier to attachment limits the rate of creation of reconstructed
islands. At high fluxes, most of the material present at the surface is
non-transformed (c). Data for (1) $F$$=$10$^{-4}$BL$\cdot$s$^{-1}$, (2)
$F$$=$10$^{-3}$BL$\cdot$s$^{-1}$, and (3) $F$$=$10$^{-2}$BL$\cdot$s$^{-1}$
are shown.}
\label{kin}
\end{figure}

The presence of the barriers to attachment in the model
introduces specific features into its dynamical behavior.
In the dependence of $N$ on the total coverage $\Theta_{\rm tot}$
(Fig.~\ref{kin}a) the nucleation onset (fast increase of $N$)
takes place at higher coverages than in standard (minimal)
growth model (Fig.~\ref{kin}b), Ref.~\onlinecite{commin}.
The maxima of $N(\Theta_{\rm tot})$ are sharp and occur at higher
$\Theta_{\rm tot}$ than the broad maxima of the standard model.

Island growth in our model starts by a transformation of 3
neighboring HUCs.\cite{tochi1} Due to the barriers to attachment this
happens only after a certain time elapsed from HUC clustering .
This time does {\it not\/} scale with flux
in contrast to the time for clustering of units of material.
Transformation becomes the rate-limiting process of island formation.
By the time the transformation begins, more material has been deposited
for the case of higher flux $F$.
This shifts the nucleation onset and $N_{\rm max}(\Theta_{\rm tot})$
position to higher $\Theta_{\rm tot}$. With increasing $F$,
amount of non-transformed material at a given $\Theta_{\rm tot}$
increases (Fig.~\ref{kin}c), and so does the disorder of
the quenched sample.

\begin{figure}
\epsfxsize=6.3cm
\centerline{\epsfbox{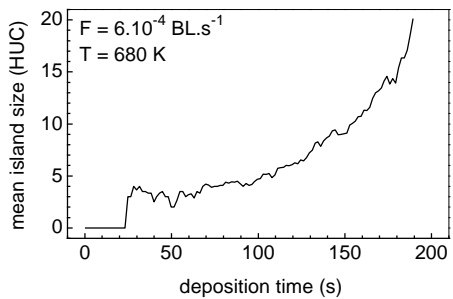}}
\caption{The delay of island creation is clearly visible in the record
of the mean island size time evolution. Instability of newborn
islands causes the decrease of $\langle s\rangle$ at short times.}
\label{mean}
\end{figure}

A~newborn island composed of 3 units can further grow or decay.
The competition between growth and decay of transformed clusters
results in a steep decrease of
$N(\Theta_{\rm tot})$ for $N>N_{\rm max}$.
This decrease is {\it not\/} due to coalescence, because the
islands at $N$$=$$N_{\rm max}$ are small.
Both the delay in island formation and the instability of islands
leave traces in the time evolution of the mean island size,
cf. Fig.~\ref{mean}.

The shift of $N_{\rm max}$ to higher $\Theta_{\rm tot}$ than in
standard model and the fast decrease of $N(\Theta_{\rm tot})$
after reaching $N_{\rm max}$ contribute to
the high value of $\chi$ observed at $\Theta_{\rm tot}$=const
during growth with barriers to attachment.  With the barrier
to attachment decreasing and the nearest-neighbor
bond strength increasing, $\chi$ decreases.

Analytical theories usually relate the value of the growth
exponent $\chi$ to $i^{\ast}$, the number of material units
in a ``critical'' (i.e., largest unstable) island in the
growth system.   For the determination of
$i^{\ast}$, we can use a formula $\chi=2i^{\ast}/(i^{\ast}\!+3)$
derived by Kandel.\cite{kandel,podkandel}
$\chi$ in the model
varies smoothly with $E_{S}$, $E_N^X$, and $E_N^T$
so that the corresponding values of $i^{\ast}$ are
non-integer numbers between $i^{\ast}$=1 ($\chi$=0.5) and
$i^{\ast}$=2 ($\chi$=0.8). In addition, a dependence of
the island density on $E_{S}$ (the substrate contribution
to the hopping barrier) was found, in contrast to
predictions of Ref.~\onlinecite{thib}.

\begin{figure}
\epsfxsize=7.5cm
\centerline{\epsfbox{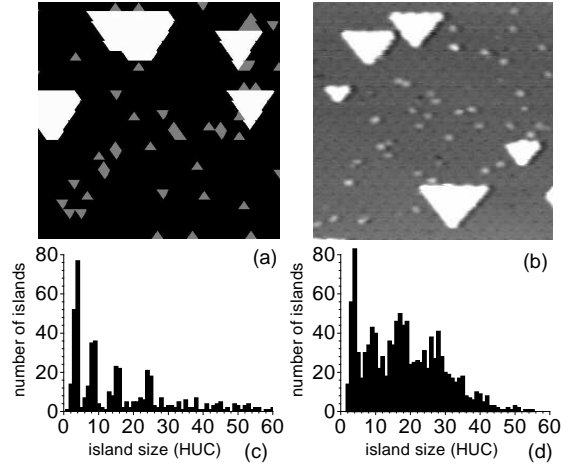}}
\caption{Growth morphologies in the model (a) and
experiment (b). During growth, magic island sizes are
stabilized, resulting in a non-trivial
island size distribution (c, d), Ref.~\protect\onlinecite{magic}.}
\label{morph}
\end{figure}

The instability of growing islands does not strongly
affect morphologies obtained for Si/Si(111)7$\times$7
MBE growth.  We observe the characteristic
multiple-peaked island size distribution (Fig.~\ref{morph}c, d),
but with broader peaks (in better agreement with the
experimental results) as compared to the model with
detachment of material from islands forbidden.\cite{magic}
In the morphologies of experimental samples, non-transformed
clusters of Si adatoms formed during quenching of samples
are visible
(Fig.~\ref{morph}b). The density of non-transformed material
may be experimentally determined and
compared to the model results.

{\it Equilibration.\/} The decay of Si/Si(111)7$\times$7 reconstructed
1~BL-high adatom (A) and vacancy (V) islands was experimentally
studied in Refs.~\onlinecite{dece,dece1,eqsi}.  Using STM at an elevated
temperature, the authors of Refs.~\onlinecite{dece} and \onlinecite{dece1}
followed a number of isolated
(a nearest island or a step edge at a distance more than
800~\AA) adatom or vacancy islands and studied the
temperature dependence of their decay rates.
The decay rates $\nu^{A}$, $\nu^{V}$ showed Arrhenius behavior
$\nu^{A,V}$$=$$\nu_{0}^{A,V}\exp(-E_{a}^{A,V}/k_{B}T)$ with
$\nu_{0}^{A}$$=$2$\times$10$^{11\pm 1}$adatoms$\cdot$s$^{-1}$,
$E_{a}^{A}$$=$1.5$\pm$0.1~eV for adatoms,
$\nu_{0}^{V}$$=$3$\times$10$^{9\pm 1}$adatoms$\cdot$s$^{-1}$,
$E_{a}^{V}$$=$1.3$\pm$0.2~eV for vacancies, respectively.
The decay rate of vacancy islands was
found to be approx. 5 times  lower than that of adatom
islands.\cite{diff}

With our model, we traced the
evolution of a 96-HUC compact adatom or vacancy island placed
on a vicinal Si surface (U-type steps, the terrace width of 480~nm, the
distance from the descending step edge of 240~nm) equilibrated
at a given temperature. Step edges on the vicinal surface form
adatom
sources and traps necessary for true disappearance of a single
adatom or vacancy island in a model with
periodic boundary conditions.

\begin{figure}
\epsfxsize=6.3cm
\centerline{\epsfbox{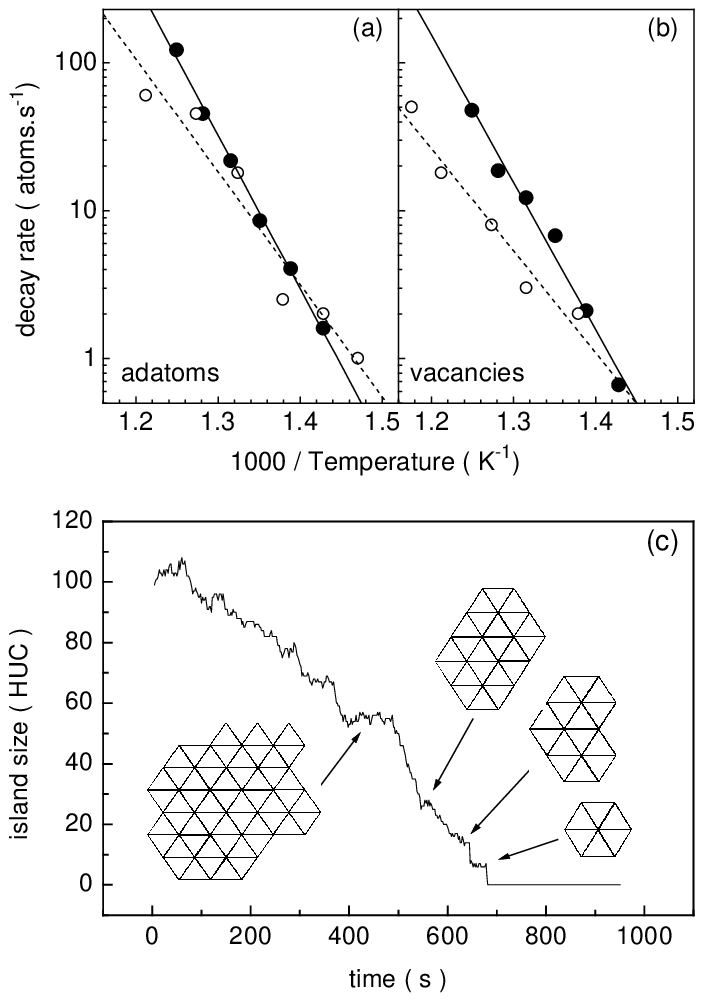}}
\caption{Arrhenius plots of experimental ($\circ$)
(Ref.~\protect\onlinecite{dece}) and
modeled ($\bullet$) decay rates for adatom (a) and vacancy (b) islands.
In the model, the rate of vacancy filling is 2$\times$ lower than that
of adatom islands decay.
During island decay, ``magic'' island sizes are stabilized (c). Insets
show some of the observed stable island morphologies. These are close
to equilibrium island shapes\protect\cite{dece,dece1,eqsi} and  differ from
``magic'' island shapes observed during growth
(cf.~Fig~\protect\ref{morph}b).}
\label{decay}
\end{figure}

The temperature dependence of the decay rates for adatom
and vacancy islands in our model is shown in
Fig.~\ref{decay}a, b. With the parameters listed above,
the decay rates in the model are higher than the experimental
ones ($\nu_{0}^{A}$$=$3$\times$10$^{14.5\pm
0.5}$adatoms$\cdot$s$^{-1}$, $E_{a}^{A}$$=$2.1$\pm$0.1~eV,
$\nu_{0}^{V}$$=$1$\times$10$^{14\pm 1}$adatoms$\cdot$s$^{-1}$,
$E_{a}^{V}$$=$2.0$\pm$0.1~eV), and the decay rate of vacancy islands
is approx. 2 times lower than the decay rate of adatom islands.

The authors of Ref.~\onlinecite{dece,dece1} attributed the difference
between the decay rates of adatom and
vacancy islands to the effect of the Ehrlich-Schwoebel (step-edge)
barrier in the Si/Si(111)7$\times$7 system. We do not believe that
the Ehrlich-Schwoebel barrier plays any role:
Growth experiments provide no compelling evidence of the presence
of an appreciable Ehrlich-Schwoebel barrier at step edges
on Si/Si(111)7$\times$7 surface
within the relevant temperature range.\cite{xcomp2}

We also modeled adatom and vacancy islands decay with
the barrier to attachment ``switched off''.\cite{commin} The
decay rates thus obtained were lower
($\nu_{0}^{A}$$=$5$\times$10$^{13\pm 1}$adatoms$\cdot$s$^{-1}$,
$E_{a}^{A}$$=$2.1$\pm$0.2~eV,
$\nu_{0}^{V}$$=$1$\times$10$^{12\pm 1}$adatoms$\cdot$s$^{-1}$,
$E_{a}^{V}$$=$2.1$\pm$0.2~eV), but the decay rate of vacancy islands was
still approx. 2~times lower than for adatom islands.
This observation agrees with results of a standard growth model on
square lattice.\cite{natori}
The difference in these decay rates on the
vicinal surface thus seems to originate from the difference
of geometry of adatom and vacancy island boundaries.\cite{natori}

In Fig.~\ref{decay}c, a typical time evolution of the size of
a decaying island in a model with the barriers to attachment is
shown. We see that stable
(``magic'') Si islands do exist.  They correspond to equilibrium
island shapes experimentally observed \cite{dece,dece1,eqsi} and differ
from magic shapes observed during Si/Si(111)7$\times$7 growth.
\cite{magic} Magic islands are compact
(2 nearest-neighbors for all perimeter HUCs) and the barrier
to attachment prevents their shape from ``being spoiled''
by attachment of material surrounding the island. No stable
shapes are observed during island decay for the model
without barriers to attachment.

In this work, we presented a coarse-grained model of
Si/Si(111)7$\times$7 MBE growth with an activation barrier
to attachment of new material to existing islands implemented.
We demonstrated that this barrier
contributes to the steep growth-rate dependence of the island
density  observed in
Si/Si(111)7$\times$7 MBE and helps to stabilize ``magic''
island shapes in both growth and
relaxation experiments.

This work was supported by the Grant Agency of the Czech Republic,
project GA\v{C}R~202/97/1109.

\narrowtext

\end{multicols}


\begin{references}
\bibitem{commin} A~result obtained using our model with a
zero barrier to transformation is presented.
\bibitem{diff}
The difference in the adatom and vacancy island decay rates\cite{dece1}
resulted from a correction of raw data for STM tip effects.
Without this correction, the decay rates were equal within the
experimental error:
$\nu_{0}^{A}$$=$2$\times$10$^{10\pm 1}$adatoms$\cdot$s$^{-1}$,
$E_{a}^{A}$$=$1.3$\pm$0.1~eV,
$\nu_{0}^{V}$$=$4$\times$10$^{9\pm 1}$adatoms$\cdot$s$^{-1}$,
$E_{A}^{V}$$=$1.2$\pm$0.2~eV.
\end{references}
\end{document}